\documentclass[12pt]{article}
\usepackage{graphicx,epsf}

\newsavebox{\sboxpubnumber}
\newsavebox{\sboxpubdate}
\newcommand{\pubdate}[1]{\begin{lrbox}{\sboxpubdate}{#1}\end{lrbox}}
\newcommand{\pubnumber}[1]{\begin{lrbox}{\sboxpubnumber}{\begin{tabular}{l} #1 \\
				 \usebox{\sboxpubdate}
				 \end{tabular}}
                           \end{lrbox}
                           \pubblock}
\newcommand{\Title}[1]{\begin{center} {\Large #1 } \end{center}}
\newcommand{\Author}[1]{\begin{center}{ \sc #1} \end{center}}
\newcommand{\Address}[1]{\begin{center}{ \it #1} \end{center}}

\newcommand{\pubblock}{\rightline{
			\usebox{\sboxpubnumber}}}
\newenvironment{Abstract}{\begin{quotation}  }{\end{quotation}}
\newenvironment{Presented}{\begin{quotation} \begin{center}
             PRESENTED AT\end{center}\bigskip
      \begin{center}\begin{large}}{\end{large}\end{center}
      \end{quotation}}

\begin{document}

%%%%%%%%%%%%%%%%%%%%%%%%%%%%%%%%%%%%%%%%%%%%%%%%%%%%%%%%%%%%%%%%%%%%%%%%
%%
%% START EDITING HERE!
%%
%%%%%%%%%%%%%%%%%%%%%%%%%%%%%%%%%%%%%%%%%%%%%%%%%%%%%%%%%%%%%%%%%%%%%%%%
\begin{titlepage}
\pubdate{\today}                    %fill in the date
\pubnumber{UCLA/01/TEP/34} %preprint number(s)

\vfill
\Title{Baryogenesis in the wake of inflation}
\vfill
\Author{Alexander Kusenko
%\footnote{And possible funding acknowledgements.
%                           DELETE THIS FOOTNOTE IF UNNECESSARY!!}
}
\Address{Department of Physics and Astronomy, UCLA, Los Angeles, CA
90095-1547 \\
and \\
RIKEN BNL Research Center, Brookhaven National
Laboratory, Upton, NY 11973
}
\vfill

\begin{Abstract}

Electroweak baryogenesis could be very efficient at the end of an
electroweak-scale inflation.  Reheating that followed inflation could
create a highly non-equilibrium plasma, in which the baryon number
violating transitions were rapid.  In addition, the time-dependent motions
of the scalar degrees of freedom could provide the requisite CP violation.
If the final reheat temperature was below 100~GeV, there was no wash-out of
the baryon asymmetry after thermalization.  The observed value of the
baryon asymmetry can be attained in a number of models, some of which do
not require a significant departure from the Standard Model.

\end{Abstract}
\vfill
\begin{Presented}
    COSMO-01 \\
    Rovaniemi, Finland, \\
    August 29 -- September 4, 2001
\end{Presented}
\vfill
\end{titlepage}
\def\thefootnote{\fnsymbol{footnote}}
\setcounter{footnote}{0}

%%%%%%%%%%%%%%%%%%%%%%%%%%%%%%%%%%%%%%%%%%%%%%%%%%%%%%%%%%%%%%%%%%%%%%%%
% The document starts here
%%%%%%%%%%%%%%%%%%%%%%%%%%%%%%%%%%%%%%%%%%%%%%%%%%%%%%%%%%%%%%%%%%%%%%%%
\section{Introduction}

Inflation probably took place in the early universe.  If 
the scale of (the latest) inflation was of the order of the electroweak
scale, the subsequent reheating would provide ideal conditions for
electroweak baryogenesis.

The original scenario of Kuzmin, Rubakov, and Shaposhnikov~\cite{krs,rs}
was based on a brilliant idea that all three Sakharov's
conditions~\cite{sakharov} necessary for a successful baryogenesis were
satisfied, at least qualitatively, in the early universe due to the
properties of the Standard Model at finite temperature.  First, the baryon
number was violated by sphaleron transitions.  Second, the universe was out
of thermal equilibrium during a phase transition.  Finally, CP was broken, 
because it is not an exact symmetry in the Standard Model.

Unfortunately, this scenario, in its original form, cannot account for the
observed baryon-to-photon ratio $\eta \sim 10^{-10}$.  First problem has to
do with the fact that the electroweak phase transition is not sufficiently
strongly first-order unless the Higgs mass is below about 45~GeV, which is
ruled out by experiment.  Second, the CP violation from the 
Cabibbo-Kobayashi-Maskawa matrix is way too small for baryogenesis because
its contribution to $\eta$ is suppressed by high powers of the Yukawa
couplings.

At the end of inflation, however, the universe was emphatically out of
equilibrium.  Could the baryon number violating processes take place in
that non-thermal environment?  The answer is yes; this was clearly
demonstrated by a consensus of numerical~\cite{ggks} and
analytical~\cite{kt,ggks,ck} arguments.  Finally, the requisite CP
violation may come from time-dependent solutions for the scalar zero-modes
in the background~\cite{ck}.  Altogether, electroweak baryogenesis can be
very efficient at the end of inflation. 

\section{Electroweak baryogenesis at preheating}

The idea~\cite{kt,ggks} of using for baryogenesis the non-equilibrium nature
of preheating~\cite{kls}, along with the new sources of CP violation that
preheating offers~\cite{ck}, is very appealing because
it is simple and requires only a minor ``upgrade'' of the Standard Model.
The inflaton has to be added to the Standard Model, but this is a widely
accepted piece of new physics that is desirable for other reasons.  The
only uncomfortable features of the scenario proposed independently in
Refs.~\cite{kt} and \cite{ggks} are the relatively low scale of inflation
and the need to couple the inflaton to the Higgs sector.  In combination
with the constraint that inflation must last for more than 60 e-folds, this
forces the inflaton potential to be unnaturally flat.  

Of course, if the universe has undergone more than one period of inflation,
the latest, electroweak-scale inflation could have lasted for only a small
number of e-folds.  In this case, there is no problem with naturalness. 

Low-scale inflation
models have been constructed~\cite{hybrid}.  One of the side benefits of
lowering the inflation scale is avoiding the gravitino over-production
constraints. There are other particle-physics motivations for using the TeV
scale, which is associated with supersymmetry breaking in a class of
models~\cite{gauge}.

Although the simplest models of hybrid
inflation do not yield parametric resonance~\cite{tachion}, there is no
fundamental reason why a weak-scale inflation could not be followed by
resonant preheating~\cite{linde}.  The latter possibility allows for
efficient electroweak baryogenesis at preheating. 

\subsection{Baryon number non-conservation during preheating}

\begin{figure}[htb]
    \centering
\hspace*{-5.5mm}
\leavevmode\epsfysize=6.5cm \epsfbox{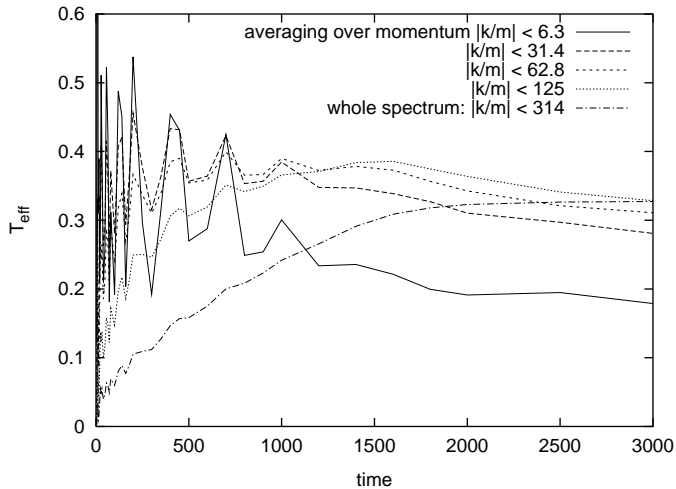}\\[3mm]
    \caption{Evolution of the effective temperature with time in a numerical
    simulation of electroweak baryogenesis at preheating~\cite{ggks}}
    \label{fig:1}
\end{figure}

\begin{figure}[htb]
    \centering
\hspace*{-5.5mm}
\leavevmode\epsfysize=6.5cm \epsfbox{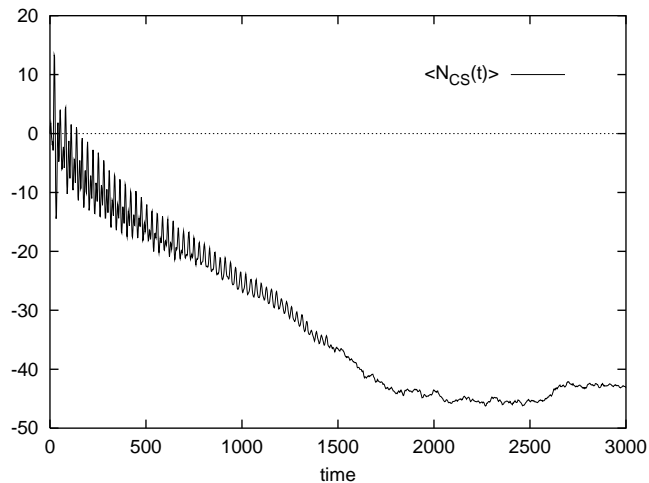}\\[3mm]
    \caption{
Change in Chern-Simons number during the run shown in Fig.~1
}
    \label{fig:2}
\end{figure}

Numerical simulations in 1+1 dimensions~\cite{ggks,num} demonstrate that
the baryon number violating processes can go unsuppressed in a non-thermal
plasma at preheating.  One way to think about this is in terms of
``effective temperature''~\cite{ggks}.  Although plasma is far from
equilibrium, the Chern-Simons number changes at the same rate as in thermal
plasma at temperature $T_{\rm eff}$.  Since sphalerons, being extended
objects, are mainly affected by the long-wavelength modes, it is the
infra-red part of the spectrum that determines $T_{\rm eff}$.  At
preheating, the long-wavelength modes are overpopulated as compared to the
thermal ensemble.  Hence, the effective temperature $T_{\rm eff}$ is higher
than the energy density.  When thermal equilibrium is restored, the final
temperature $T_R < 100~{\rm GeV}<T_{\rm eff}$.  At that point the rate of
baryon number violating processes is negligible.  This behavior, favorable
to baryogenesis, is shown in Figs.~1,2.

One can try to describe this behavior semi-analytically by isolating
certain gauge degrees of freedom and studying their evolution at
preheating~\cite{ck}, when the vector boson mass, $M_{W}$ changes with time
due to the oscillations of the inflaton coupled to the Higgs sector.  Let us
introduce an {\em ansatz}~\cite{bc,co89} for the gauge potential 
\begin{equation}   %2
gA_{\mu}=(\frac{\tau_a}{2i})A_{\mu}^a.
\end{equation}   
Our spatially-homogeneous {\it ansatz} is:
\begin{equation}   %3
gA_0=0;\;gA_i=(\frac{\tau_i}{2i})\phi(t)
\end{equation}
in which the group index is tied to the spatial index.  By the conventional
rules of charge conjugation and parity for the gauge potential, $\phi$ is C
even, P odd, CP odd.

It is important to note that this {\it ansatz} does not correspond to a
non-vanishing VEV for an EW field.  Gauge invariance alone is enough to
ensure that there can be no expectation value coupling the space-time
indices to group indices.

One readily calculates the EW electric and magnetic fields:
\begin{equation}   %4 
gE_i\equiv G_{0i}=(\frac{\tau_i}{2i})\dot{\phi}(t);\;
gB_i\equiv \frac{1}{2}\epsilon_{ijk}G_{jk}=(\frac{\tau_i}{2i})\phi^2.
\end{equation}
Then one calculates the density $W$ of the Chern-Simons number as:
\begin{equation}     %5
W=(\frac{1}{8\pi^2})\phi^3.
\end{equation}
It is straightforward to check that $\dot{W}$ is the topological charge
density $Q$, related to B+L violation through the anomaly equation.

With the assumption of a given Higgs VEV, the equations of motion for the
gauge potential are:
\begin{equation}     %6
[D^{\mu},G_{\mu\nu}]+M_W^2(t)(A_{\nu}+(\partial_{\nu}U)U^{-1})=0.
\end{equation}
Here the unitary matrix $U$ represents the Goldstone (phase) part of the
Higgs field.  The mass term will be assumed to have the form:
\begin{equation}     %7
M_W^2(t)=m^2(1+\epsilon \cos (\omega t))
\end{equation}
where $m$ is the value of $M_W$ with no oscillations.  

This {\em ansatz} has some important properties.  First, it carries some
non-zero Chern-Simons density.  Second, the equation of motion for the
field $\phi$,
\begin{equation}     %8
\ddot{\phi}+2\phi^3+(1+\epsilon \cos r t)\phi =0,
\end{equation} 
has resonantly growing solutions.  Finally, the ansatz can evolve for a
long time without breaking spatial homogeneity~\cite{ck}.  
\begin{figure}[htb]
    \centering
\hspace*{-5.5mm} \leavevmode\epsfysize=6.5cm \epsfbox{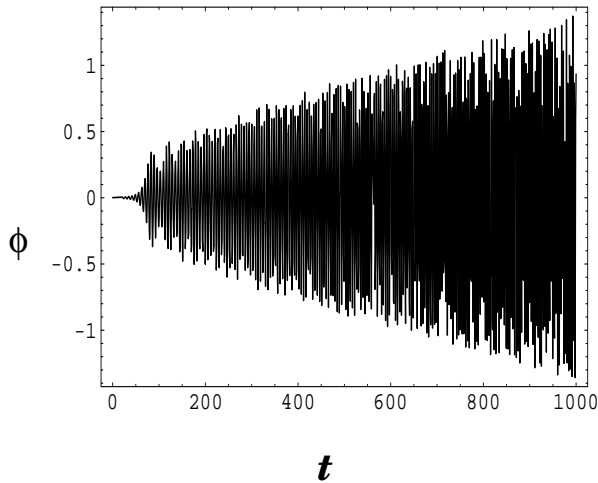 
}\\[3mm]
\caption{ Resonant growth of the Chern-Simons number density in the gauge
``condensate'' for the case of slowly changing driving frequency~\cite{ck}.
} \label{fig:3}
\end{figure}

The Chern-Simons density grows as the third power of $\phi$.  

\subsection{CP violation at preheating} 

CP and time-reversal symmetries at preheating may be broken by the
time-dependent classical motions of scalar fields in the
background~\cite{cgk}.  This is similar to the source of CP violation used
by Affleck and Dine~\cite{ad}.  CP violation of this kind is poorly
constrained by experiment.  Several viable scenarios for electroweak
baryogenesis at preheating were presented in Ref.~\cite{cgk}.  For example,
a modified spontaneous baryogenesis a la Cohen, Kaplan, and
Nelson~\cite{ckn} becomes very efficient at preheating.  Their original
scenario used the variation of the Higgs field inside a wall of a bubble
formed in a first-order phase transition.  A similar effect can occur at
preheating uniformly in space, on the horizon scales~\cite{cgk}.  One can
get the desired baryon asymmetry in a Standard Model supplemented by an
additional Higgs doublet and an inflaton sector~\cite{cgk}.  The difference
with the scenario proposed by Cohen, Kaplan and Nelson~\cite{ckn} is that
in our case CP violation occurs homogeneously in space, as opposed to in a
bubble wall. In addition, the final prediction for baryon asymmetry in the
CKN scenario was very far from the equilibrium value  because the
sphaleron rate was slow on the time scales associated with the growth of
the bubble.  In our case, the Higgs parameters change slowly in time while
the baryon number non-conservation is rapid.  This allows a slow adiabatic
adjustment of the baryon number to that which minimizes the free energy.

\section{Other possibilities for baryogenesis at preheating}

Non-thermal production of heavy particles at the end of
inflation~\cite{wimpzillas} could result in baryogenesis~\cite{GUT}, or a
leptogenesis~\cite{lepto} via the decay of the heavy particles.  

Alternatively, in models with low-energy supersymmetry, weak-scale
inflation can lead to formation of the Affleck-Dine
condensate~\cite{ad,drt}.  In general, this condensate does not remain
homogeneous and can break up~\cite{ks} into SUSY Q-balls~\cite{ak_mssm}.
This affords a number of interesting possibilities for generating baryons
and dark matter simultaneously~\cite{ks,em,kk,ls}.

If there were complex scalar fields not associated with supersymmetry,
their fragmentation in a similar fashion could produce stable Q-balls that
could make up the dark matter.  Interestingly enough, this form of dark
matter can be strongly self-interacting~\cite{k_st,e_self} and can,
therefore, resolve the problems with non-interacting cold dark matter on
small scales~\cite{ss}.

%%%%%%%%%%%%%%%%%%%%%%%%%%%%%%%%%%%%%%%%%%%%%%%%%%%%%%%%%%%%%%%%%%%%%%%%
%%
%%   use this format to include an .eps figure into your paper
%%
%\begin{figure}[htb]
%    \centering
%    \includegraphics[height=1.5in]{cosmo.eps}
%    \caption{Figure caption.}
%    \label{fig:cosmo}
%\end{figure}
%%%%%%%%%%%%%%%%%%%%%%%%%%%%%%%%%%%%%%%%%%%%%%%%%%%%%%%%%%%%%%%%%%%%%%%%

\section{Conclusion}

A highly non-thermal state of the universe at the end of inflation provides
a fertile ground for baryogenesis.  In addition to a dramatic departure
from thermal equilibrium, preheating creates conditions for baryon number
non-conservation and activates sources of CP violation that are largely
unconstrained by the experimental data.  Several modest modifications of the
Standard Model allow for efficient electroweak baryogenesis in the wake of  
inflation.


\begin{thebibliography}{99}

%%
%%  bibliographic items can be constructed using the LaTeX format in
%%  SPIRES:
%%  see    http://www.slac.stanford.edu/spires/hep/latex.html
%%  SPIRES will also supply the CITATION line information; please
%%  include it.
%%


\bibitem{krs}  V.~A.~Kuzmin, V.~A.~Rubakov, and M.~Shaposhnikov,
Phys. Lett. {\bf 155B}, 36 (1985).  
%%CITATION = PHLTA,B155,36;%% 

\bibitem{rs} V.~A.~Rubakov and M.~E.~Shaposhnikov,
%``Electroweak baryon number non-conservation in the early universe and in
%high-energy collisions,'' 
Usp.\ Fiz.\ Nauk {\bf 166}, 493 (1996)
[Phys.\ Usp.\  {\bf 39}, 461 (1996)];
%%CITATION = HEP-PH 9603208;%%  

\bibitem{sakharov} A. D. Sakharov, JETP Lett. {\bf 6}, 24 (1967).

\bibitem{kuzmin} V.~A.~Kuzmin, Pisma ZhETP, {\bf 13}, 335 (1970).  

\bibitem{kt}  L.~M.~Krauss and M.~Trodden, Phys. Rev. Lett.{\bf 83}, 1502
(1999). 
%%CITATION = HEP-PH 9902420;%%

\bibitem{ggks}  J.~Garc\'ia-Bellido, D.~Grigoriev, A.~Kusenko, and
M.~Shaposhnikov, Phys. Rev. D {\bf 60}, 123504 (1999).  
%%CITATION = HEP-PH 9902449;%%

\bibitem{ck}  J.~M.~Cornwall and A.~Kusenko,
%``Baryon number non-conservation and phase transitions at preheating,''
Phys.\ Rev.\ D {\bf 61}, 103510 (2000)
%%CITATION = HEP-PH 0001058;%%

\bibitem{cgk} J.~M.~Cornwall, D.~Grigoriev and A.~Kusenko,
%``Resonant amplification of electroweak baryogenesis at preheating,''
Phys.\ Rev.\ D {\bf 64}, 123518 (2001)
%%CITATION = HEP-PH 0106127;%%

\bibitem{kls}  L.~Kofman, A.~Linde, and A.~A.~Starobinsky,
Phys. Rev. Lett. {\bf 73}, 3195 (1994), {\it ibid.} {\bf 76}, 1011 (1996);
Phys. Rev. D {\bf 56}, 3258 (1997).  
%%CITATION = HEP-TH 9405187;%%
%%CITATION = HEP-TH 9510119;%%
%%CITATION = HEP-PH 9704452;%%


\bibitem{hybrid} A.D. Linde, Phys. Lett. {\bf B259}, 38 (1991); Phys.
Rev. D {\bf 49}, 748 (1994); L.~Knox and M.~Turner, Phys. Rev. Lett. {\bf
70} (1993) 371; L. Randall, M. Solja\v ci\'c, and A. H. Guth, Nucl.
  Phys. B {\bf 472}, 377 (1996); J. Garc{\'\i}a-Bellido, A. D. Linde
  and D. Wands, Phys.  Rev. D {\bf 54}, 6040 (1996); 
%%CITATION = ASTRO-PH 9209006;%%
%%CITATION = HEP-PH 9512439;%%
%%CITATION = ASTRO-PH 9605094;%%
D.~H.~Lyth and E.~D.~Stewart,
%``More varieties of hybrid inflation,''
Phys.\ Rev.\ D {\bf 54}, 7186 (1996); 
%%CITATION = HEP-PH 9606412;%
J.~Garc{\'\i}a-Bellido, these Proceedings.  

\bibitem{gauge} G.~F.~Giudice and R.~Rattazzi,
%``Theories with gauge-mediated supersymmetry breaking,''
Phys.\ Rept.\  {\bf 322}, 419 (1999). 
%%CITATION = HEP-PH 9801271;%%

\bibitem{tachion} G.~N.~Felder, J.~Garcia-Bellido, P.~B.~Greene, L.~Kofman,
A.~D.~Linde and I.~Tkachev,
%``Dynamics of symmetry breaking and tachyonic preheating,''
Phys.\ Rev.\ Lett.\  {\bf 87}, 011601 (2001).
%%CITATION = HEP-PH 0012142;%%

\bibitem{linde} A. Linde, private communications.

\bibitem{num}
J.~Garcia-Bellido and D.~Y.~Grigoriev,
%``Inflaton-induced sphaleron transitions,''
JHEP {\bf 0001}, 017 (2000).
%%CITATION = HEP-PH 9912515;%%
A.~Rajantie, P.~M.~Saffin and E.~J.~Copeland,
%``Numerical simulations of electroweak baryogenesis at preheating,''
arXiv:hep-ph/0010347; 
%%CITATION = HEP-PH 0010347;%%
E.~J.~Copeland, D.~Lyth, A.~Rajantie and M.~Trodden,
%``Hybrid inflation and baryogenesis at the TeV scale,''
Phys.\ Rev.\ D {\bf 64}, 043506 (2001); 
%%CITATION = HEP-PH 0103231;%%
A.~Rajantie, these Proceedings, hep-ph/0111200.
%%CITATION = HEP-PH 0111200;%%

\bibitem{bc} K.~M.~Bitar and S.~Chang,
%``Vacuum Tunneling Of Gauge Theory In Minkowski Space,''
Phys.\ Rev.\  {\bf D17}, 486 (1978).
%%CITATION = PHRVA,D17,486;%%

\bibitem{co89} J.~M.~Cornwall,
%``High Temperature Sphalerons,''
Phys.\ Rev.\  {\bf D40}, 4130 (1989).
%%CITATION = PHRVA,D40,4130;%%

\bibitem{ad} 
I.~Affleck and M.~Dine,
%``A New Mechanism For Baryogenesis,''
Nucl.\ Phys.\ B {\bf 249}, 361 (1985).
%%CITATION = NUPHA,B249,361;%%

\bibitem{ckn} A.~G.~Cohen, D.~B.~Kaplan and A.~E.~Nelson,
%``Spontaneous baryogenesis at the weak phase transition,''
Phys.\ Lett.\ B {\bf 263}, 86 (1991).
%%CITATION = PHLTA,B263,86;%%

\bibitem{wimpzillas} D.J.~Chung, E.W.~Kolb, and A.~Riotto, Phys. Rev. Lett.
{\bf 81}, 4048 (1998); Phys. Rev. {\bf D59}, 023501 (1999). 
%%CITATION = HEP-PH 9802238;%%

\bibitem{GUT}
E.~W.~Kolb, A.~D.~Linde and A.~Riotto,
%``GUT baryogenesis after preheating,''
Phys.\ Rev.\ Lett.\  {\bf 77}, 4290 (1996).
%%CITATION = HEP-PH 9606260;%%

\bibitem{lepto}
G.~F.~Giudice, M.~Peloso, A.~Riotto and I.~Tkachev,
%``Production of massive fermions at preheating and leptogenesis,''
JHEP {\bf 9908}, 014 (1999).
%%CITATION = HEP-PH 9905242;%%

\bibitem{drt} M.~Dine, L.~Randall and S.~Thomas, Nucl. Phys. {\bf B458}
  (1996) 291; 
A.~Anisimov and M.~Dine,
%``Some issues in flat direction baryogenesis,''
arXiv:hep-ph/0008058.
%%CITATION = HEP-PH 0008058;%

\bibitem{ks} 
A.~Kusenko, M.~Shaposhnikov:
%``Supersymmetric Q-balls as dark matter,''
Phys.\ Lett.\  B {\bf 418}, 46 (1998)
%%CITATION = HEP-PH 9709492;%%

\bibitem{ak_mssm} A.~Kusenko,
%``Solitons in the supersymmetric extensions of the standard model,''
Phys.\ Lett.\ B {\bf 405}, 108 (1997); 
%%CITATION = HEP-PH 9704273;%%
Phys.\ Lett.\ B {\bf 404}, 285 (1997);  
%%CITATION = HEP-TH 9704073;%%
A.~Kusenko, V.~Kuzmin, M.~E.~Shaposhnikov and P.~G.~Tinyakov,
%``Experimental signatures of supersymmetric dark-matter Q-balls,''
Phys.\ Rev.\ Lett.\  {\bf 80}, 3185 (1998).
%%CITATION = HEP-PH 9712212;%%

\bibitem{em} K.~Enqvist, J.~McDonald: 
%``Q-balls and baryogenesis in the MSSM,''
Phys.\ Lett.\  B {\bf 425}, 309 (1998); 
%%CITATION = HEP-PH 9711514;%%
%``B-ball baryogenesis and the baryon to dark matter ratio,''
Nucl.\ Phys.\  B {\bf 538}, 321 (1999);
%%CITATION = HEP-PH 9803380;%%
%``D-term inflation and B-ball baryogenesis,''
Phys.\ Rev.\ Lett.\  {\bf 81}, 3071 (1998);
%%CITATION = HEP-PH 9806213;%%
%``MSSM dark matter constraints and decaying B-balls,''
Phys.\ Lett.\  B {\bf 440}, 59 (1998);
%%CITATION = HEP-PH 9807269;%%
%K.~Enqvist and J.~McDonald,
%``Observable isocurvature fluctuations from the Affleck-Dine condensate,''
Phys.\ Rev.\ Lett.\  {\bf 83}, 2510 (1999)
%%CITATION = HEP-PH 9811412;%%
%``The dynamics of Affleck-Dine condensate collapse,''
%%CITATION = HEP-PH 9908316;%%

\bibitem{kk} 
S.~Kasuya and M.~Kawasaki,
%``Q-ball formation: Obstacle to Affleck-Dine baryogenesis in the
%gauge-mediated SUSY breaking?,''
Phys.\ Rev.\ D {\bf 61}, 041301 (2000);
%%CITATION = HEP-PH 9909509;%%
Phys.\ Rev.\ D {\bf 62}, 023512 (2000); 
%%CITATION = HEP-PH 0002285;%%
Phys.\ Rev.\ D {\bf 64}, 123515 (2001)
[arXiv:hep-ph/0106119].
%%CITATION = HEP-PH 0106119;%%
S.~Kasuya,
%``Difficulty of a spinning complex scalar field to be dark energy,''
Phys.\ Lett.\ B {\bf 515}, 121 (2001)
[arXiv:astro-ph/0105408].
%%CITATION = ASTRO-PH 0105408;%%

\bibitem{ls} 
M.~Laine and M.~E.~Shaposhnikov,
%``Thermodynamics of non-topological solitons,''
Nucl.\ Phys.\ B {\bf 532}, 376 (1998)
[arXiv:hep-ph/9804237].
%%CITATION = HEP-PH 9804237;%%

\bibitem{k_st} A.~Kusenko and P.~J.~Steinhardt,
%``Q-ball candidates for self-interacting dark matter,''
Phys.\ Rev.\ Lett.\  {\bf 87}, 141301 (2001)
[arXiv:astro-ph/0106008].
%%CITATION = ASTRO-PH 0106008;%%

\bibitem{e_self} 
K.~Enqvist, A.~Jokinen, T.~Multamaki and I.~Vilja,
%``Constraints on self-interacting Q-ball dark matter,''
arXiv:hep-ph/0111348.
%%CITATION = HEP-PH 0111348;%%

\bibitem{ss} 
D.~N.~Spergel and P.~J.~Steinhardt,
%``Observational evidence for self-interacting cold dark matter,''
Phys.\ Rev.\ Lett.\  {\bf 84}, 3760 (2000); 
%%CITATION = ASTRO-PH 9909386;%%
P.~Steinhardt, these Proceedings. 

\end{thebibliography}
\end{document}